\documentclass{aa}  
\usepackage{amsmath}
\usepackage{tensor}
\usepackage{graphicx}
\renewcommand{\phi}{\varphi}

\usepackage{todonotes}

\usepackage[normalem]{ulem}

\begin{document}

\title{Superdiffusion of energetic particles at shocks: A L\'evy Flight model for acceleration}
\titlerunning{Superdiffusive acceleration of energetic particles at shocks}

\author{Sophie~Aerdker\inst{1,2}
\and Lukas~Merten\inst{1,2}
\and Frederic~Effenberger\inst{1,2}
\and Horst~Fichtner\inst{1,2}
\and Julia~Becker Tjus\inst{1,2,3}}
\authorrunning{Aerdker et al.}


\institute{ Ruhr-Universit\"at Bochum, Fakult\"at f\"ur Physik und Astronomie, Institut f\"ur Theoretische Physik IV, Universit\"atsstra\ss e 150, 44780 Bochum, Germany \and Ruhr Astroparticle and Plasma Physics Center (RAPP Center), Bochum, Germany \and Chalmers University of Technology, Department of Space, Earth and Environment,  412 96 Gothenburg, Sweden}

\abstract
{In the Heliosphere, power-law particle distributions are observed e.g.\ upstream of interplanetary shocks, which can result from superdiffusive transport. This non-Gaussian transport regime may result from intermittent magnetic field structures. Recently, we showed that a L\'evy flight model reproduces the observed features at shocks: power-law distributions upstream and enhanced intensities at the shock.}
{We extend the L\'evy flight model to study the impact of superdiffusive transport on particle acceleration at shocks. The acceleration time scale and spectral slope are compared to Gaussian diffusion and a L\'evy walk model. }
{The fractional transport equation is solved by sampling the number density with the corresponding stochastic differential equation that is driven by an alpha-stable L\'evy distribution. For both Gaussian and superdiffusive transport we use a modified version of CRPropa~3.2.}
{We obtain the number density and energy spectra for constant and energy-dependent anomalous diffusion and find, compared to the case of Gaussian diffusion, harder energy spectra at the shock as well as faster acceleration. The spectral slope is even harder than predicted for L\'evy walks.}
{ L\'evy flight models of superdiffusive transport lead to observed features in the Heliosphere. We further show that superdiffusive transport impacts the acceleration process by changing the probability to escape the shock. The flexibility of the L\'evy flight model allows for further studies in the future, taking the shock geometry and magnetic field structure into account. }

\maketitle

\section{Introduction}

Cosmic rays (CRs) undergo random scattering due to magnetic field fluctuations present in astrophysical plasmas. When successive small-angle scattering leads to Brownian motion of the particles this process can be described by Gaussian spatial diffusion along the magnetic field \citep{Kulsrud-Pearce-69, Skilling-1975}. The time evolution of the particles' distribution function may then be described by the transport equation (e.g.~\citet{Parker-1965, Schlickeiser2002}) which has been successfully applied to model Galactic CR transport (see e.g.~ \citet{BeckerTjusMerten2020, Mertsch2020} for reviews).

Observations in the Heliosphere, however, indicate that particle transport may not be Gaussian but \emph{anomalous} \citep{Perri-etal-2022}. Such evidence comes from power-law profiles of ion and electron fluxes upstream of interplanetary shocks \citep{Perri-Zimbardo-2007, Perri-Zimbardo-2008jgr, Giacalone-2012, Perri-etal-2022}, at the solar wind termination shock \citep{Perri-Zimbardo-2009, Perri-Zimbardo-2012}, or solar energetic particle events \citep{Trotta-Zimbardo-2011}, which contradicts expectations for Gaussian diffusion. Anomalous diffusion may result from intermittent magnetic fields: Particles scatter in magnetic field structures but move freely between them \citep{ShukurovEA2017, Luebke-etal-2024}.

Recently, we showed how superdiffusive transport of energetic particles at shocks leads to the observed upstream power-law distributions and intensity peaks at the shock \citep{Effenberger-etal-2024}. A fractional diffusion and a L\'evy flight model were introduced to describe superdiffusive transport. 

In this paper, we extend the L\'evy flight model from \cite{Effenberger-etal-2024} to include the acceleration of particles at the shock. The expected energy spectrum and acceleration time scale give further insights into superdiffusion and its role in particle transport. 
We compare the energy spectra we obtain with the L\'evy flight model to those obtained by standard diffusive shock acceleration and by a L\'evy walk model as discussed in \citet{Perri-Zimbardo-2012}.

In general, the term anomalous diffusion describes all dynamics for which the mean squared displacement is characterized by a non-linear dependence on time,
\begin{equation}
\label{eq:msqd}
\langle(\Delta x)^2\rangle \propto \tilde{\kappa}_\zeta t^\zeta,
\end{equation}
in contrast to normal diffusion ($\zeta = 1$). Depending on the anomalous diffusion exponent $\zeta$, it is distinguished between subdiffusive ($\zeta < 1$) and superdiffusive processes ($\zeta > 1$). Usually $\zeta \leq 2$ is the upper limit for superdiffusive transport, with $\zeta = 2$ corresponding to ballistic transport or free streaming. Note, that the anomalous diffusion coefficient $\tilde{\kappa}_{\zeta}$ has units $\mathrm{m}^2/\mathrm{s}^{\zeta}$. 

Anomalous transport cannot be captured by the standard transport equation but can be described by \emph{fractional} transport equations. Fractional transport equations can, e.g., be solved by applying the generalized It\^o lemma to obtain the corresponding stochastic differential equations (SDEs) that are driven by an alpha-stable L\'evy distribution \citep{Ito51, Magdziarz-Weron-2007}. Section \ref{sec:fractionaldiffusion} gives an overview of fractional transport equation and the L\'evy flight model we employ to solve for the particle number density.

The L\'evy flight model is more flexible than, e.g.\ Fourier series approximations \citep{Stern-etal-2014, Effenberger-etal-2024}: The energy gain at the shock can be calculated along with the particle transport, the (anomalous) diffusion coefficient can be momentum-dependent and, 2D or 3D transport along magnetic field lines can all be modeled with a similar simulation set-up, which is not possible with the given Fourier series approximation. We present first results for particle acceleration at a 1D planar shock, contrasting constant and momentum-dependent anomalous diffusion coefficients. 

For Gaussian diffusion, the energy spectrum of particles at a stationary non-relativistic planar shock with compression ratio $q$ has the characteristic slope $s = 2 - 3q/(q-1)$, see, e.g., \citet{Drury}. Even for non-linear diffusion this \emph{universal power-law} can be found \citep{WalterEA2022}. Considering L\'evy flights, we find harder\footnote{corresponding to smaller values for the power-law slope parameter $p^-s$, here $s > -2$. More particles gained high energies.} energy spectra at the shock, similar to what \citet{Perri-Zimbardo-2012} find for L\'evy walks. 

We compare the time-dependent spectra and number densities at the shock to those obtained with constant Gaussian diffusion as well as with a time-dependent Gaussian diffusion coefficient that mimics the mean squared displacement of the corresponding anomalous diffusion exponent $\zeta$. This allows to differentiate between effects coming from the larger mean-squared displacement of the L\'evy flight model at a given time and those coming from the different underlying stochastic processes. 

\section{Fractional Diffusion and L\'evy Flights}
\label{sec:fractionaldiffusion}
 
\begin{figure}
\includegraphics[width=0.5\textwidth]{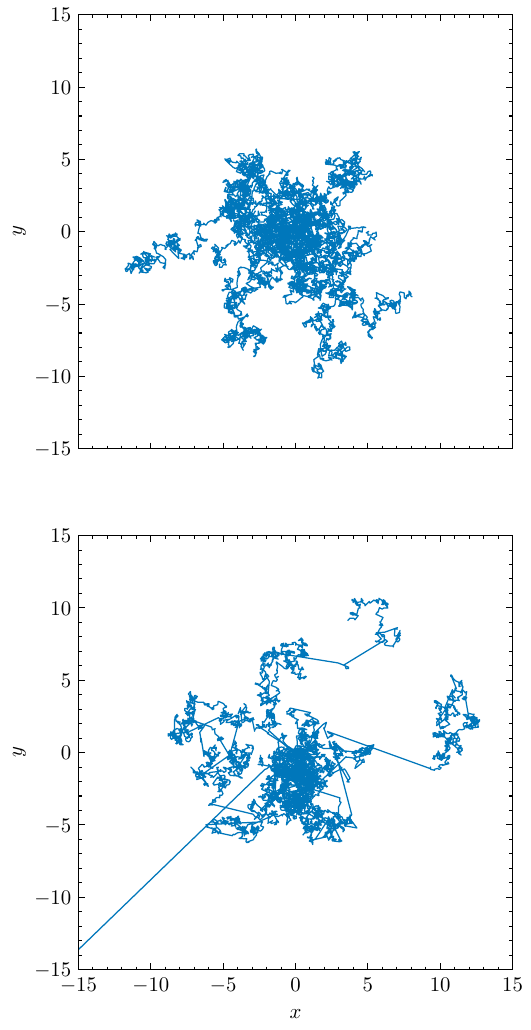}
\caption{Ten pseudo-particle trajectories starting at $(0,0)$ for a Gaussian diffusion process ($\alpha = 2.0$, with $\kappa_{2} = 1$ top) and L\'evy process ($\alpha = 1.7$, $\kappa_{1.7} = 1$, bottom). We discuss how to obtain pseudo-particle trajectories and their meaning in Sec.~\ref{sec:pseudo-particle-trajectories}. }
\label{fig:particle-orbits}
\end{figure}

A common way to model superdiffusive processes is the usage of L\'evy flights. Like Brownian motion, L\'evy flights are Markov processes but the jump length distribution follows an inverse power-law \citep{Metzler-Klafter-2000}. From the jump length distribution and the assumption of finite characteristic waiting time between such jumps a space-fractional diffusion equation can be obtained (see e.g. \citet{Metzler-Klafter-2004} and references therein):
\begin{equation}
\label{eq:diff}
\frac{\partial f(x,t)}{\partial t} =  \kappa_{\alpha} \nabla^\alpha f(x,t) \,,
\end{equation}
with the fractal dimension $\alpha$ and fractional diffusion tensor $\hat{\kappa}_{\alpha}$. The Riesz-derivative \citep{Gorenflo-etal-1999} is given by
\begin{equation}
  \label{eq:riesz-riemann}
    \nabla^{\alpha} f(x) = -\frac{1}{2\cos(\alpha\pi /2)} (\tensor*[_{-\infty}]{D}{_x^{\alpha}} + \tensor*[_x]{D}{_{+\infty}^{\alpha}}) f(x) \,,
\end{equation}
with the Riemann-Liouville fractional derivative defined as
\begin{equation}
\label{eq:riemann-liouville}
   \tensor*[_0]{D}{_t^{1-\beta}} f(t) = \frac{1}{\Gamma(\beta)}\frac{d}{dt}\int_0^t(t-s)^{\beta-1}f(s)ds \,.
\end{equation}
For $\alpha = 2$ the normal diffusion equation is recovered. Figure \ref{fig:particle-orbits} shows trajectories of Brownian motion $\alpha = 2$ and $\alpha = 1.7$ with the characteristic L\'evy flights in 2D.

The power-law distribution of the jump length leads to a diverging variance, since at a given time $t$ an arbitrarily large distance $l$ may be traveled. However, \citet{Metzler-Klafter-2000} define a 'pseudo' mean squared displacement $\left[ x^2 \right] \propto t^{2/\alpha}$ to relate the fractional dimension $\alpha$ to the anomalous diffusion exponent $\zeta = 2/\alpha$. Note, that the fractional diffusion coefficient $\kappa_{\alpha}$ has units $\mathrm{m}^{\alpha}/\mathrm{s}$. For $\alpha = 2$ eq.~(\ref{eq:diff}) reproduces Gaussian diffusion. 

Still, the diverging mean squared displacement is an issue for the spatial transport of massive particles. This can be solved by using L\'evy walks instead. By coupling the jump probability of a distance $l$ to the time $t$ that is needed to travel such a distance, a well-defined mean squared displacement is recovered \citep{Metzler-Klafter-2000, Metzler-Klafter-2004, Zaburdaev-etal-2015}. L\'evy walks have already been used to model particle transport and acceleration at shocks \citep{Perri-Zimbardo-2012}. The spatio-temporal coupling leads to a different dynamical process, which cannot easily be modeled with our approach. However, L\'evy flights are a good approximation (we refer to \citet{Perri-etal-2015} for a comparison of a L\'evy flight and  L\'evy walk model). Thus, we investigate how superdiffusive acceleration is different when applying a L\'evy flight model compared to L\'evy walks.

In order to investigate superdiffusive shock acceleration with L\'evy flights, we solve the one-dimensional fractional transport equation assuming isotropic particle distributions in momentum space. The time evolution of the differential number density $\mathcal{N} = p^2 f(x, p, t)$ in space $x$ and momentum $p$, with $f$ being the particle distribution function, is determined by
\begin{align}
\label{eq:TransportEq}
    \frac{\partial \mathcal{N}}{\partial t} &= \kappa_{\alpha} \nabla^{\alpha} \mathcal{N}  - u(x) \cdot \frac{\partial  \mathcal{N}}{\partial x} - \frac{p}{3} \frac{\partial u}{\partial x} \frac{\partial  \mathcal{N}}{\partial p} +  S(x, p, t),
\end{align}
where the spatial diffusion is generalized to superdiffusion and we consider a spatially constant anomalous diffusion coefficient $\kappa_{\alpha}$. In this macroscopic picture, (super-)diffusive shock acceleration results from adiabatic heating due to the divergence of the background flow $u(x)$ (e.g.\ \citet{Krymskii77, KirkSchlickeiser88, Drury}). Sources and sinks are given by $S(x,p,t)$.

\subsection{Solving the Fractional Fokker-Planck Equation with Stochastic Differential Equations}
\label{sec:SDE}

The fractional transport equation (\ref{eq:TransportEq}) is difficult to solve analytically, due to the non-local Riesz-derivative. When energy changes are neglected, the solution of the fractional diffusion equation or fractional diffusion-advection equation can e.g. be approximated by Fourier Series \citep{Stern-etal-2014, Effenberger-etal-2024}. A more flexible method is to sample the solution with a Monte-Carlo approach. 

For that, the transport equation (\ref{eq:TransportEq}) is written into a fractional Fokker-Planck equation (FFPE) which corresponds to a set of SDEs according to It\^o calculus \citep{Ito51}. In order to model one-dimensional superdiffusive shock acceleration, diffusion in momentum is neglected and the diffusion coefficient $\kappa_\alpha$ is considered to be constant in space. The spatial displacement is then described by the SDE
\begin{equation}
\label{eq:levy-SDE}
\mathrm{d}x(t) = u \,\mathrm{d}t + \sqrt{2}\kappa_\alpha^{1/\alpha}\,\mathrm{d}L_\alpha(t) \,.
\end{equation}
Here $\mathrm{d}L_\alpha(t)$ is an $\alpha$-stable L\'evy process, for $\alpha = 2$ is identical to a Wiener process, and results in Brownian motion. The first term describes the deterministic motion due to advection, the second term the stochastic motion, characterized by the fractional diffusion coefficient $\kappa_{\alpha}$ and the L\'evy process $\mathrm{d}L_\alpha(t)$. 

The change in momentum is given by an ordinary differential equation since diffusion in momentum space is neglected, 
\begin{equation}
\label{eq:ode-momentum}
\mathrm{d}p(t) = - \frac{p}{3} \frac{\partial u}{\partial x} \,\mathrm{d}t \;.
\end{equation}
Note that eq.~(\ref{eq:levy-SDE}) and (\ref{eq:ode-momentum}) do not describe the time evolution of actual particles, but sample the differential number density $\mathcal{N}$. 

SDEs can be approximated numerically in their integral form (see e.g.\ \citet{KloedenPlaten92, Gardiner2009}). A modified version of CRPropa~3.2 \citep{CRPropa3.2} is used to solve the stochastic differential equation in space which uses the Euler-Maruyama scheme for integrating the SDE (\ref{eq:levy-SDE})
\begin{equation}
\label{eq:EMScheme}
    x_{n+1}-x_n = u \Delta t + \sqrt{2} \kappa_{\alpha}^{1/\alpha} \Delta t^{1/\alpha} \eta_{\alpha, t}\;.
\end{equation}
 In each simulation time step, $\Delta t$, a random number $\eta_{\alpha,t}$ is drawn from the $\alpha$-stable L\'evy distribution, and the pseudo-particles position is updated to $x_{n+1}$. The random number generation for $\eta_{\alpha,t}$ is based on the Chambers-Mallows-Stuck algorithm \citep{Chambers-etal-1976}. L\'evy flights result from random numbers drawn from the heavy power-law tails of the distribution. For $\alpha = 2$ the Euler-Maruyama scheme with a Wiener process is recovered.
 
 In general, the Euler-Maruyama scheme has no constraints on the time step. However, the choice of time step can be crucial to obtaining correct results, especially when simulating diffusive shock acceleration. This is discussed in detail in e.g.~\citet{KruellsAchterberg94}, \citet{StraussEffenberger2017}, and \citet{Aerdker-etal-2024}. Furthermore, to enhance statistics at high energies which only a fraction of pseudo-particles ever reach, the \texttt{CandidateSplitting} module of CRPropa is used. Pseudo-particles that cross specified boundaries in energy are split into two copies, in the later analysis and they are weighted accordingly to obtain the correct spectra. Boundaries are chosen depending on the expected spectral slope to balance the number of pseudo-particles in each energy bin (see \citet{Aerdker-etal-2024} for details of this method). 

 The SDE approach is quite flexible and easy to extend to higher dimensions and other geometries compared to Fourier series approximations. In CRPropa, Eq.~(\ref{eq:EMScheme}) is defined in the lab frame. However, the diffusive step can be calculated in the orthonormal base of the magnetic field by integrating along the magnetic field line for parallel diffusion, calculating the perpendicular diffusive step, and transforming back to the lab frame in each time step (see \cite{Merten-etal-2017} for details of the transformation). 

The ordinary differential equation (\ref{eq:ode-momentum}) describing the time evolution of the pseudo-particles' momentum is integrated with an Euler scheme.

CRPropa 3.2 does not allow for continuous injection of pseudo-particles during simulation time. Instead, all pseudo-particle positions in phase-space are stored at times $T_i = i \Delta T$ during simulation. After simulation, the differential number density $n(x,p,T)$ at time $T$ is obtained by summing over all contributions $n(x,p,T_i)$ weighted by the time interval $\Delta T$. Note, that $\Delta T$ does not necessarily need to be the same as the simulation time step $\Delta t$. We refer to \citep{Merten-etal-2018, Aerdker-etal-2024} for more details. This method was already applied to model superdiffusive transport and compared to a Fourier series approximation in previous work \citep{Effenberger-etal-2024}.

\subsection{Solving the fractional diffusion equation: 2D pseudo-particle trajectories}
\label{sec:pseudo-particle-trajectories}

As a first application, we obtain the pseudo-particle trajectories for Gaussian diffusion and superdiffusion in 2D shown in Fig.~\ref{fig:particle-orbits} by setting $u = 0$. 
 
In order to model the process in 2D we make the following changes to the Euler-Maruyama scheme (\ref{eq:EMScheme}): The scattering direction is randomly chosen within $[0, 2\pi)$ and the diffusive step length $\Delta r_{\mathrm{diff}} = \sqrt{\Delta x_{\mathrm{diff}}^2 +\Delta y_{\mathrm{diff}}^2}$ is calculated in each time step. Note, that for this two random numbers are drawn from the L\'evy distribution, with 

\begin{align}
    \label{eq:2Dtrajetory}
    \Delta r_{\mathrm{diff}} = \sqrt{2} \kappa_{\alpha}^{1/\alpha} \sqrt{\eta_x^2 + \eta_y^2} \, t^{1/\alpha},
\end{align}

assuming that the diffusion coefficient $\kappa_{\alpha}$ is the same in $x$ and $y$ direction and with L\'evy random numbers $\eta_{\alpha,x}, \eta_{\alpha, y}$. This ensures a positive step length and for $\alpha = 2$ is equal to a random number drawn from the two-dimensional Chi distribution. This approach can easily be extended to three dimensions.

\section{Modeling Superdiffusive Shock Acceleration}
\label{sec:SSA}

 For the superdiffusive shock acceleration (SSA) with L\'evy walks, \citet{Perri-Zimbardo-2012} found that the energy spectrum is slightly harder than predicted for the normal diffusive shock acceleration (DSA). For L\'evy flights, we thus expect the spectral slope to be flatter as well, but not necessarily to be same as for L\'evy walks. 

\citet{WalterEA2022} found that the spectral slope does not differ from Gaussian diffusive shock acceleration for non-linear diffusion where the diffusion coefficient depends on the distribution function. Also, for time-dependent diffusion we do not expect a different spectral slope. However, similar to findings by \citet{WalterEA2022} the acceleration time scale is affected by the increasing diffusion coefficient over time. 

In the macroscopic picture of the transport equation, acceleration at the shock arises from the interplay between (anomalous) diffusion and advection. We briefly explain constraints on modeling DSA or SSA with SDEs. We then compare the results for SSA with L\'evy flights to time-dependent diffusion coefficient modeling a process with the same (pseudo) mean squared displacement. 

\subsection{One-dimensional planar shock}

 In the diffusive picture, momentum gain at shocks is described by eq.~(\ref{eq:ode-momentum}) so that the advective speed $u(x)$ must be continuously differentiable. Thus, a finite shock width is assumed instead of a discrete shock transition. The one-dimensional planar shock at $x = 0$ is described by
\begin{align}
\label{eq:1Dadvection}
    u(x) = \frac{u_1 + u_2}{2} - \frac{u_1 - u_2}{2} \tanh\left( \frac{x}{L_{\mathrm{sh}}}\right), 
\end{align}
 with upstream and downstream velocity, $u_1$ and $u_2 = u_1/q$, with compression ratio $q$, and shock width $L_{\mathrm{sh}}$. The velocity profile $u(x)$ is also used in other studies of DSA \citep{KruellsAchterberg94, AchterbergSchure2010, WalterEA2022, Aerdker-etal-2024}. 

In the following, all units are normalized so that
\begin{equation}
    \tilde{x} = \frac{x}{x_0},\quad \tilde{u} = \frac{u}{u_0}, \quad \tilde{t} = \frac{t}{t_0}, \quad \tilde{\kappa} = \frac{\kappa}{\kappa_0},\quad \tilde{p} =  \frac{p}{p_0},
\end{equation} 
with $x_0/v_0 = t_0$ and $p_0$ being the momentum of pseudo-particles injected at the shock. For modeling transport and acceleration in the Heliosphere, the normalization can be set e.g. to $x_0 = 1\,\mathrm{AU}$ and $v_0 = 400\,\mathrm{km/s}$. The compression ratio is assumed to be $q = 4$.

\subsubsection*{Constraints on the simulation time step}

The finite shock width in eq.~(\ref{eq:1Dadvection}), leads to constraints on the simulation time step and diffusion coefficient when modeling an ideal shock (for discussion on normal diffusion, see e.g. \citet{KruellsAchterberg94, AchterbergSchure2010, Aerdker-etal-2024}). From a numerical perspective, pseudo-particles need to encounter the changing advection to be accelerated, which implies a sufficiently small time step. However, with small time steps, the diffusive step $\Delta x_{\mathrm{diff}} = \sqrt{2} \kappa_{\alpha}^{1/\alpha} \Delta t^{1/\alpha}$ --- a measure for the stochastic step --- may become smaller than the shock width. In that case, pseudo-particles will not make it back to the shock to be repeatedly accelerated. This constraint essentially depends on the chosen diffusion coefficient and shock width. 

To model an ideal shock, the parameter $\epsilon = u_1 L_{\mathrm{sh}}/\kappa_{2}$ should be lower than one for Gaussian diffusion. Note that the diffusion coefficient has units $\mathrm{m}^\alpha / \mathrm{s}$ when L\'evy flights are considered. We found, however, that the constraints in the time step are less restrictive for superdiffusive transport since occasionally L\'evy flights make it possible to cross the shock front again, even if the mean diffusive step is small. 

\subsubsection*{Simulation results}

The time-dependent number density and momentum spectrum at the shock with superdiffusion, $\alpha = 1.7$, are compared to normal diffusion, $\alpha = 2$, and normal diffusion with a time-dependent diffusion coefficient that mimics the same dependency of the mean squared displacement on time, $\kappa_{2}(t) = t^{2/\alpha-1}$ with $\alpha = 1.7$. 

\begin{figure*}
    \centering
    \includegraphics[width = 1\textwidth]{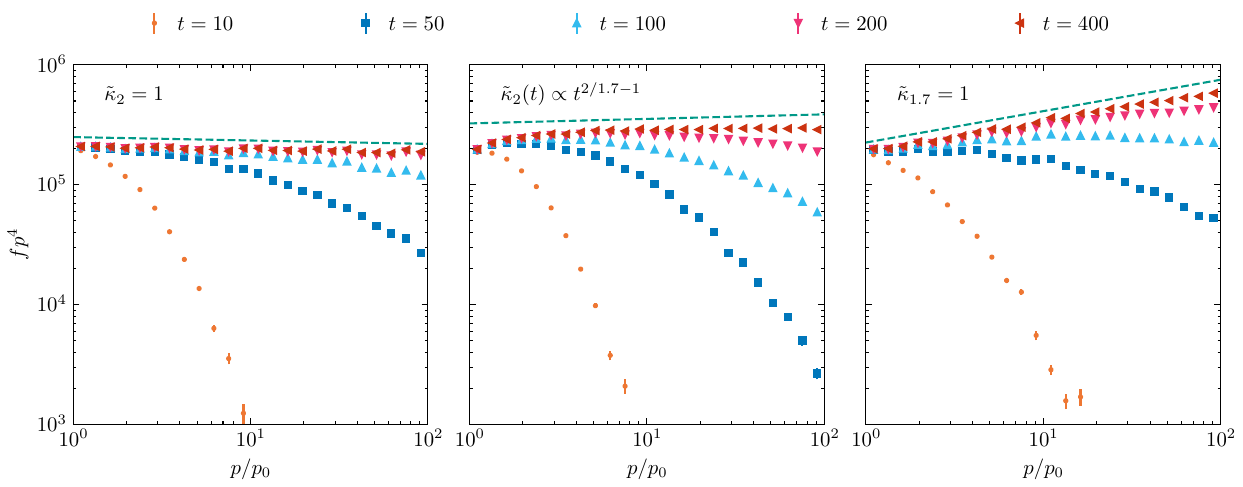}
    \caption{Time evolution of the weighted spectrum $f(p,t)p^4$ at the shock ($\tilde{x} = [0,1]$) for normal diffusion, $\tilde{\kappa}_2 = 1$ (left), time-dependent diffusion, $\tilde{\kappa}_2(t) \propto \tilde{t}^{2/1.7-1}$ (middle) and L\'evy flights with $\alpha = 1.7$ and $\tilde{\kappa}_{1.7} = 1$ (right). Time-dependent diffusion and L\'evy flights with $\alpha = 1.7$ have the same (pseudo) mean squared displacement. The dashed lines indicate the slope of the fitted spectra at time $\tilde{t} = 800$.}
     
    \label{fig:const-spectra}
    
\end{figure*}

Figure \ref{fig:const-spectra} shows the time evolution of the spectra $fp^2$ at the shock ($x = [0,1]$) weighted by $p^2$ for the different diffusion processes/diffusion coefficients. The dotted lines show the fitted slope of the spectra when the stationary solution for $p <= 10^2\,p_0$ is reached. The spectra are fitted in the range $p = [10, 10^2]\,p_0$ to exclude effects from pseudo-particle injection at $p = p_0$.

In the case of a constant Gaussian diffusion coefficient (left), the stationary spectrum has the expected slope $-2.017 \pm 0.003$\footnote{The spectra are fitted for $p < 10^2\,p_0$ at $\tilde{t} = 800$ by a least squares method.} which is already reached at $\tilde{t} = 200$. Acceleration at the shock with a constant diffusion coefficient is fast compared to that with a diffusion coefficient which increases over time (middle). The time-dependent diffusion coefficient slows down the acceleration at the shock, thus, at $\tilde{t} = 200$ the stationary solution for $p < 10^2$ is not yet reached, as expected for an increased diffusion coefficent in the estimate for the acceleration time $\tau_{acc}$, see eq. (\ref{eq:acc-time}) below. The stationary spectral slope is the same as for a constant Gaussian diffusion coefficient. The dip at low momentum (approx.\ $p < 3$) comes from the vanishing diffusion coefficient early in time ($\kappa_2(t = 0) = 0$) and constraints on the chosen shock width/simulation time step. 

The right panel of fig.~\ref{fig:const-spectra} shows the time evolution resulting from superdiffusive shock acceleration. The stationary spectrum is harder compared to normal diffusion, $s_{1.7} = 1.750 \pm 0.007$ for $p < 10^2$. Also, the stationary solution at $\tilde{t} = 200$ for $p < 10^2\,p_0$ is not yet reached. However, acceleration is more efficient than in the case of time-dependent diffusion. The results are consistent with the work of \citet{Perri-Zimbardo-2012} who also found flatter spectra for superdiffusive shock acceleration with L\'evy walks.

\begin{figure*}
    \centering
    \includegraphics[width = 1\textwidth]{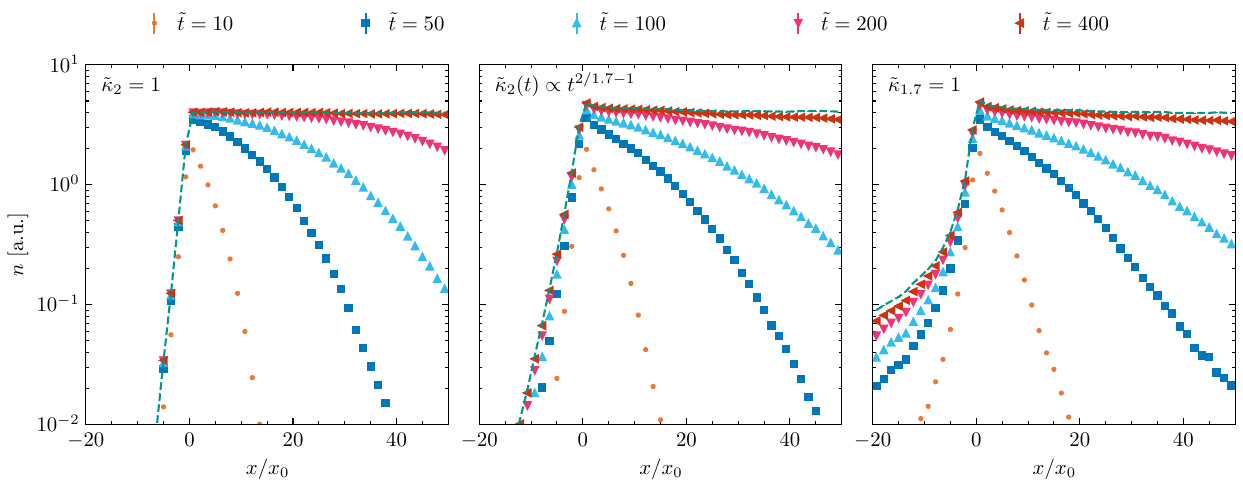}
    \caption{Time evolution of the number density integrated over momentum $n = \int \mathcal{N} \,\mathrm{d}p$ for normal diffusion, $\tilde{\kappa}_2 = 1$ (left), time-dependent diffusion, $\tilde{\kappa}_2(t) \propto \tilde{t}^{2/1.7-1}$ (middle), and L\'evy flights with $\alpha = 1.7$ and $\tilde{\kappa}_{1.7} = 1$ (right). The dashed line shows the number density at $\tilde{t} = 800$. }
    \label{fig:const-distriution}
\end{figure*}

Figure \ref{fig:const-distriution} compares the time-evolution of the differential number density integrated over momentum. With increasing Gaussian diffusion over time, more particles reach the upstream region against the background flow compared to constant Gaussian diffusion. With L\'evy flights, the characteristic power-law distributions emerge upstream. For a detailed discussion, see \citet{Effenberger-etal-2024}. Note, that the number densities differ, since here the advection speed drops at the shock. 

The dashed lines show the number density profiles at $\tilde{t} = 800$. Since there are no significant changes in the range $\tilde{x} = [-20, 50]$, the steady-state solution is already reached for Gaussian diffusion and almost for time-dependent Gaussian diffusion and L\'evy flights. 

\subsection{Comparison to L\'evy walks }

The question remains why fractional diffusion changes the universal shock spectrum in contrast to time-dependent or even non-linear diffusion coefficients. At a one-dimensional planar shock, the spectral slope only depends on the shock compression ratio and the escape probability (see, e.g., \citet{Drury}). Since the compression ratio remains the same, superdiffusion modeled with L\'evy flights must result in a lower escape probability, leading to more efficient acceleration. 

For L\'evy walks, \citet{Perri-Zimbardo-2012} obtain with the propagator approach the escape probability $P_{\mathrm{esc}}$ and with that the power-law 
\begin{equation}
    \label{eq:slope-levy-walks}
    \gamma = P_{\mathrm{esc}} \frac{E}{\Delta E} = 6 \frac{\mu - 2}{\mu - 1} \frac{1}{q - 1} + 1\;,
\end{equation}
for superdiffusive shock acceleration with $\Delta E / E$ being the relative energy gain of relativistic particles at a planar shock with compression ratio $q$. Their parameter $\mu$ can be related to the anomalous diffusion exponent $\zeta = 4 - \mu$. Thus, the fractional dimension $\alpha = 2 / (4 - \mu)$ has the same (pseudo) mean squared displacement. For the relation between the escape probability and spectral index we also refer to \citet{Kirk-etal-96}. They found softer spectra for subdiffusive shock acceleration.

\begin{figure*}
    \centering
    \includegraphics[width = 1\textwidth]{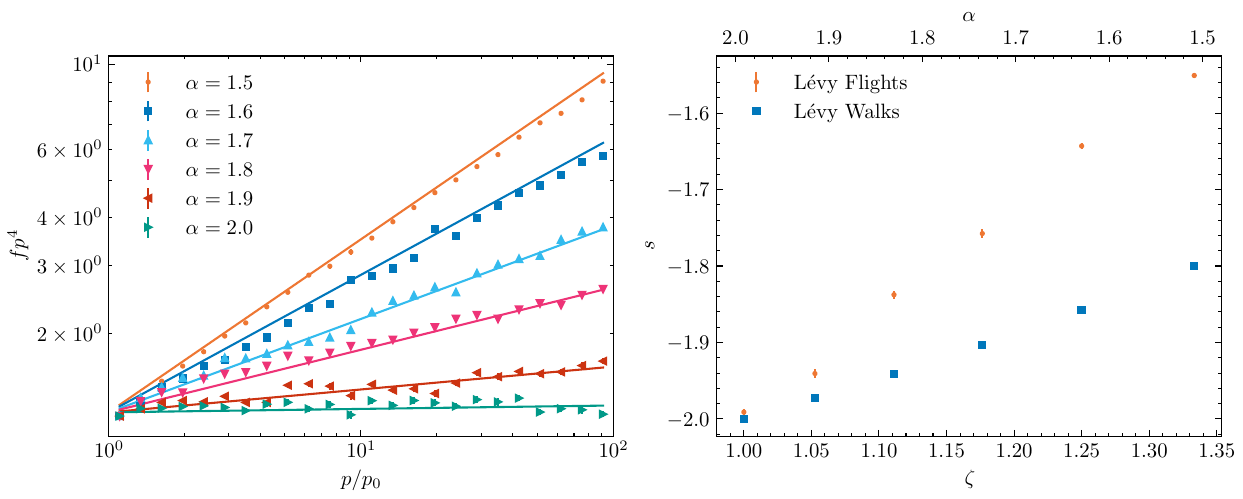}
    \caption{Stationary spectra for $p <= 100p_0$ for different fractional dimension $\alpha$ considering L\'evy flights (left). Dots are simulation results, lines indicate the slopes fitted to the range $p = (10, 100)p_0$. The fitted slopes are shown in comparison to the slopes for L\'evy walks in the right panel depending on the fractional dimension $\alpha$ and (pseudo) mean squared displacement. }
    \label{fig:spectra-levyflights-walks}
\end{figure*}

Figure \ref{fig:spectra-levyflights-walks} compares the spectra of superdiffusive shock acceleration obtained with L\'evy flights to those with L\'evy walks having the same anomalous diffusion exponent $\zeta$. We do not find the same spectral slopes for L\'evy flights but even harder spectra. Thus, the jump length distribution of L\'evy flights seems to lower the escape probability even more, which leads to more efficient acceleration. 

\subsection{Comparison of particle number densities}

\begin{figure*}
    \centering
    \includegraphics[width = 1\textwidth]{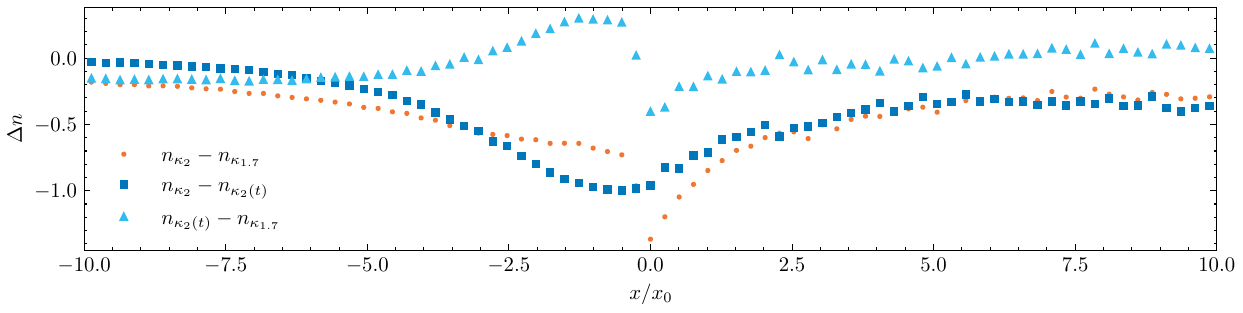}
    \caption{Difference in differential number densities comparing Gaussian diffusion and L\'evy flights (orange dots), constant and time-dependent Gaussian diffusion (blue squares), time-dependent Gaussian diffusion and L\'evy flights (light blue triangles). }
    \label{fig:ratio-plots}
\end{figure*}

Another approach to analyzing the escape probability is to have a look at the differential number densities for the different diffusion processes. Fig.~\ref{fig:ratio-plots} compares the differential number density of all considered diffusion processes in the range $\tilde{x} = [-10, 10]$.

Differences between the processes are subtle: For both, time-dependent Gaussian diffusion and L\'evy flights more particles are upstream and close to shock, indicated by the negative ratio in the upper and middle panel of Fig.~\ref{fig:ratio-plots}. For L\'evy flights, the upstream power-laws are visible as more particles make it far upstream, even compared to time-dependent Gaussian diffusion (at least, at that point in time $\tilde{t} = 800$, without an upper boundary for the time-dependent diffusion coefficient $\kappa_2(t)$). Also, the peak structure which forms at the shock for L\'evy flights is well visible at $x = 0$ in the upper panel. 

More interesting is the ratio between the diffusion processes which have the same (pseudo) mean squared displacement in the bottom panel. For L\'evy flights, the particle number density far upstream is higher, indicating a more efficient scattering from the downstream region back over the shock. Also, the number density right at the shock is higher. In the macroscopic approach, energy gain at the shock comes from the divergence of the velocity profile (see eq.~(\ref{eq:ode-momentum})). With more particles close to the shock, they may be accelerated more efficiently.

\subsection{Momentum-dependent diffusion}

A more realistic description of the diffusion coefficient includes a dependence on the particles' momentum. With the SDE approach momentum-dependent diffusion in 1D can be modeled within the same framework. In the following we show the resulting spectra at a shock, considering momentum-dependent (fractional) diffusion coefficients
\begin{equation}
    \label{eq:momentum-dependent-diffusion}
    \kappa_{\alpha}(p) = \kappa_0 \left(\frac{p}{p_0}\right)^\delta \;.
\end{equation}
 The fractional transport equation is solved analogously to the previous section with the diffusion coefficient in the SDE (\ref{eq:levy-SDE}) now being momentum dependent. 

 For Gaussian diffusion, the mean time to reach momentum $p$ depends on the momentum dependence of the diffusion coefficient \citep{Drury}. For $\alpha = 1$, the mean acceleration time is given by
\begin{equation}
\label{eq:acc-time}
    \bar{t}(p) \propto 
         \tau_{\mathrm{acc}} \left( \frac{p}{p_0} - 1\right), 
\end{equation}
with $\tau_{\mathrm{acc}} = 3/(u_1 - u_2) (\kappa_1/u_1 + \kappa_2/u_2)$. For Gaussian diffusion, the mean acceleration time we obtained for simulation agrees with the one given by eq.~(\ref{eq:acc-time}). 

\begin{figure*}
    \centering
    \includegraphics[width = 1\textwidth]{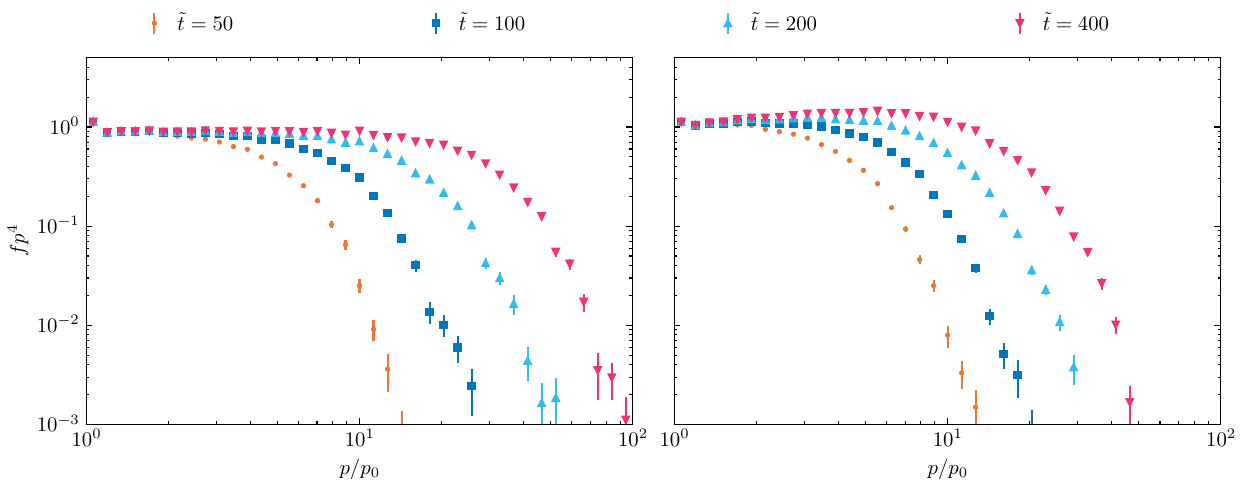}
    \caption{Time evolution of the weighted spectrum $f(p,t)p^4$ at the shock ($x = [0,1]$) for energy-dependent normal diffusion, $\kappa_2 = 1 p$ (left) and L\'evy flights with $\alpha = 1.7$ and $\kappa_{1.7} = 1 p$ (right).  }
    \label{fig:momentum-spectra}
\end{figure*}

With $\delta > 0$, momentum-dependent diffusion slows down the acceleration at the shock over time, since the particles reach higher momentum and with that the diffusion coefficient increases. Figure \ref{fig:momentum-spectra} shows the spectrum of momentum-dependent Gaussian diffusion and with L\'evy flights with $\kappa_0 = 1$ and $\delta = 1$ at the same times as for constant diffusion coefficients in Fig.~\ref{fig:const-spectra}. The cut-off due to the finite acceleration time is well visible in both cases. 

Considering superdiffusion, eq.~(\ref{eq:acc-time}) is hard to define, however, fig.~(\ref{fig:momentum-spectra}) shows that it takes longer to reach the same momentum in case of energy-dependent L\'evy flights compared to Gaussian diffusion. Note that the Gaussian diffusion and L\'evy flight process have the same value $\kappa_0 = 1$ here\footnote{Note the different physical units: $\mathrm{m^2/s}$ and $\mathrm{m^\alpha/s}$}. At a given time, the L\'evy flight model has a larger mean squared displacement which slows down the acceleration process. It is, thus, difficult to compare the different processes. 

One attempt by \citet{Perri-Zimbardo-2012} is to scale the diffusion coefficient for L\'evy walks with the fractional dimension $\mu$. This leads to a smaller anomalous diffusion coefficient and they conclude that superdiffusive acceleration is faster than Gaussian, similar to our results in the previous section. 

\begin{figure*}
    \centering
    \includegraphics[width = 1\textwidth]{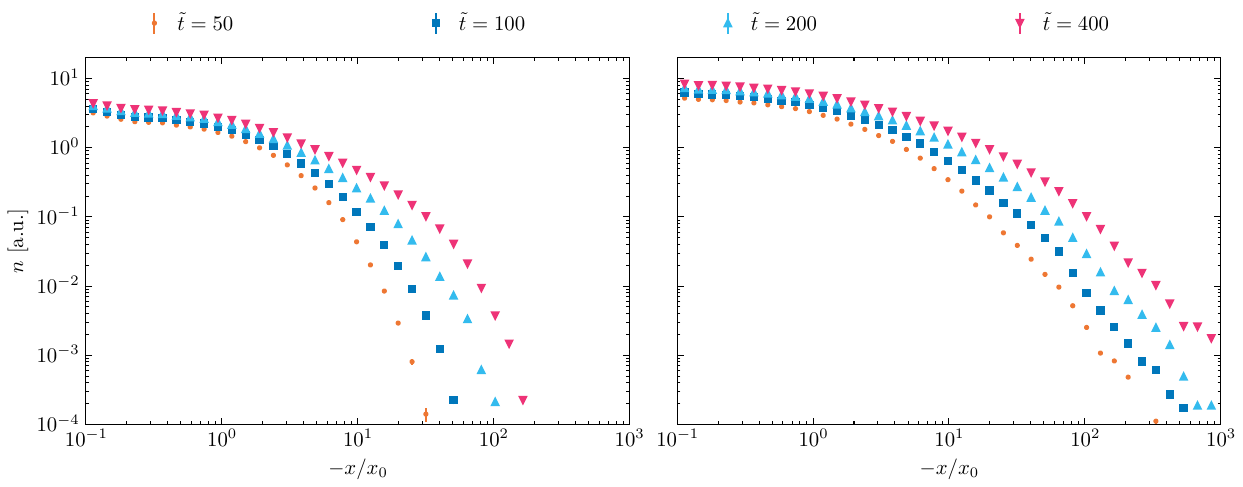}
    \caption{Time evolution of the distribution $f(x,t)$ integrated over momentum $p$ for energy-dependent normal diffusion, $\tilde{\kappa}_2 = 1 p/p_0$ (left) and L\'evy flights with $\alpha = 1.7$ and $\tilde{\kappa}_{1.7} = 1 p/p_0$ (right).  }
    \label{fig:momentum-distriution}
\end{figure*}

Over time, more particles make it to the upstream region, due to the increasing diffusion coefficient. Figure \ref{fig:momentum-distriution} shows the number density upstream of the shock over time. The upstream number density breaks from a Gaussian core to power-law tails. While the slope of the tails is determined by the fractional dimension, the break position depends on the diffusion coefficient \citep[see][]{Effenberger-etal-2024}. The effective diffusion coefficient is growing over time with particles being accelerated to higher momentum.

\section{Discussion \& Conclusions}

We presented a L\'evy flight model of superdiffusive acceleration based on stochastic differential equations which is an extension of the L\'evy flight approach described in \citet{Effenberger-etal-2024} for superdiffusive transport. The energy gain of CRs at the shock is given by Eq.~\ref{eq:ode-momentum}, which is solved along with Eq.~\ref{eq:levy-SDE} describing the L\'evy flight motion of CRs. A modified version of CRPropa3.2 is used to solve the system of (stochastic) differential equations with an Euler-Maruyama scheme. 

We find, that the energy spectra at the shock are harder when CRs are subject to L\'evy flights than for Gaussian diffusion. 
For L\'evy flights, the spectral slope depends on the fractional dimension $\alpha$ but not on the anomalous diffusion coefficient. This is analogous to Gaussian diffusion where the diffusion coefficient impacts the acceleration time scale but not the stationary spectrum. 

Energy spectra are even harder than previously found by \citet{Perri-Zimbardo-2012} for L\'evy walks. The spectral slope is determined by the compression ratio of the shock and the escape probability of the particles \citep{Drury}. With the compression ratio being the same, the probability to escape the shock must be changed due to the power-law jump length distribution of L\'evy flights. Compared to Gaussian diffusion more particles are upstream, thus, the transport back to the shock must be more efficient. Since L\'evy flights do not have a cut-off at large jumps like L\'evy walks, it is plausible that the resulting energy spectra are even harder for the same anomalous diffusion exponent $\zeta$.

The comparison of fundamentally different processes of Gaussian diffusion, L\'evy flights, and L\'evy walks is not trivial. To eliminate the effect of the increasingly higher mean squared displacement of the L\'evy flight process we further compare our results to a Gaussian process with a time-dependent diffusion coefficient. The mean squared displacement of the time-dependent Gaussian and L\'evy process are always the same, but the underlying scattering process is different. Compared to the time-dependent Gaussian diffusion, L\'evy flights have a shorter acceleration time scale. 
This is different to the approach by \citet{Perri-Zimbardo-2012} who scale down the anomalous diffusion coefficient depending on its fractional dimension for L\'evy walks to compare the acceleration time to those of Gaussian diffusion. For future work in this context, a study of the effect of so-called tempered L\'evy motion \citep{Baeumer-Meerschaert-2010} may be of interest, i.e.\ of the effect of exponentially truncated L\'evy distributions on the particle transport \citep[see, e.g.,][]{leRoux-2024}.

This work investigates anomalous diffusion at the level of particle distributions and energy spectra. 
However, it remains unclear what physical process is responsible for the non-Brownian motion. This side of the problem can e.g.~be approached by studying test particle motion in magnetohydrodynamic or synthetic turbulence with intermittent coherent structures \citep{Luebke-etal-2024}. Recently, \citet{Lemoine2023} and \citet{Kempski-etal-2023} found that CRs can be scattered on bent magnetic field lines when their gyroradius exceeds the curvature radius of the magnetic field. Such localized, strong scattering events can lead to non-Brownian diffusion on small scales \citep{Lemoine2023}. 

To make meaningful predictions on superdiffusive particle transport in astrophysical scenarios both, the fractional dimension and anomalous diffusion coefficient must be known. For Gaussian diffusion, the diffusion coefficient may be calculated from theory \citep[e.g.,][]{Shalchi-2020, Shalchi-2021}, or obtained in full-orbit test particle simulations by deriving \emph{running diffusion coefficients} $<\Delta x^2> / \Delta t$ (e.g.~\citet{Mertsch2020, ReichherzerEA2022}). 
For L\'evy flights the running anomalous diffusion coefficient would converge over $<\Delta x^2> / \Delta t^\zeta$. Thus, when the fractional dimension $\alpha =  2 / \zeta$ (or $\alpha = 4 - \zeta$ for L\'evy walks) is known, the anomalous diffusion coefficient can be determined from test particle simulations. 

The anomalous diffusion coefficient and fractional dimension can also be fitted to observations by taking into account the slope of the upstream power-law distribution (determined by the fractional dimension) and the distance from the shock at which the distribution turns into power-laws (determined by the anomalous diffusion coefficient) \citep{Effenberger-etal-2024}. 

Our work on superdiffusive transport and acceleration at 1D planar shocks sets the basis for more elaborate studies of anomalous transport. With the flexibility of the SDE approach also three-dimensional superdiffusive transport can be modeled analogously to eq.~\ref{eq:2Dtrajetory}. We already showed in Sec.~\ref{sec:pseudo-particle-trajectories} how pseudo-particle trajectories in two dimensions can be obtained, which can easily be extended to three dimensions. Also, anisotropic superdiffusion parallel and perpendicular to the magnetic field lines can be studied using the field line integration of CRPropa 3.2. Furthermore, the impact of a spherical shock geometry and the corresponding cooling due to the expanding wind can be taken into account in future analysis. 
 
\begin{acknowledgements} 
We acknowledge support from the DFG within the Collaborative Research Center SFB1491 "Cosmic Interacting Matters - From Source to Signal" (project number 445990517) and via the DFG grants EF~98/4-1 and FI~706/26-1. 
\end{acknowledgements} 

\bibliography{references}

\end{document}